\documentclass[12pt]{article}
\usepackage[dvips]{color,graphicx}
\usepackage{amsmath,amssymb}
\DeclareGraphicsExtensions{.eps}

\newcommand{\ap}{\ensuremath{\alpha'}} % Inverse string tension
\newcommand{\ls}{\ensuremath{l_s}} % String length
\def\p{\partial}

\newcommand{\tr}{\mathop{\rm Tr}}

\def\expec#1{\langle #1 \rangle}

\newcommand{\cL}{\mathcal{L}}

\newcommand{\cN}{{\mathcal{N}}}
\newcommand{\cO}{{\mathcal{O}}}

\newcommand{\bS}{{\mathbf{S}}}

\newcommand{\tret}{{t_{\mbox{\scriptsize ret}}}}
\newcommand{\tv}{{\tilde{v}}}
\newcommand{\ta}{{\tilde{a}}}

\title{\bf Acceleration and Energy Loss \\ in $\mathcal{N}=4$ SYM}
\author{Mariano Chernicoff\footnote{e-mail: mariano@nucleares.unam.mx}
~and Alberto G\"uijosa\footnote{e-mail: alberto@nucleares.unam.mx}
\\{\small Departamento de F\'{\i}sica de Altas Energ\'{\i}as,
Instituto de Ciencias Nucleares}\\ {\small Universidad Nacional
Aut\'onoma de M\'exico}\\
{\small Apdo. Postal 70-543, M\'exico D.F. 04510}}
\date{}

\begin{document}
\maketitle
\begin{abstract}
 We give a brief overview of the results obtained in \cite{chernicoff3}, concerning the rate of energy loss
 of an accelerating quark in strongly-coupled $\mathcal{N}=4$ super-Yang-Mills, both at zero and finite temperature. For phenomenological purposes, our main result is that, when a quark is created within the plasma together with its corresponding antiquark, the quark starts feeling the plasma only after the $q$-$\bar{q}$ separation becomes larger than the ($v$-dependent) screening length, and from this point on the motion is correctly described by the analytic energy loss formula previously derived by Herzog \emph{et al.} and Gubser within the stationary or late-time approximations. The present text is a slightly expanded version of two talks given at the XIII Mexican School of Particles and Fields in October 2008.\end{abstract}

%\vspace{3cm}

%\tableofcontents

\section{Introduction}

Current evidence points towards the success of the RHIC program
at recreating the long-sought quark-gluon plasma (QGP) \cite{qgprev}, scheduled to be studied in more detail at the LHC. In the range of accessible temperatures (of order a few times the deconfinement temperature), the QGP appears to behave as a strongly-coupled liquid. The phenomenon of jet quenching observed at RHIC indicates that partons crossing the QGP experience a significant energy loss \cite{energylossrev}, and so necessitates a framework for studying energy loss at strong coupling.

Motivated by this,
a considerable effort has been invested in the study of strongly-coupled thermal
non-Abelian plasmas by means of the AdS/CFT correspondence \cite{malda,gkpw,magoo}. At our present stage of knowledge, contact with the real-world  QGP produced at RHIC or LHC will be feasible only if QCD can be reasonably well approximated by at least
one of the various `QCD-like' gauge theories whose dual description is known. In the past few years, encouraging signs in this direction have emerged even for the most rudimentary
example \cite{malda}, $SU(N)$ $\mathcal{N}=4$ super-Yang-Mills (SYM), which at zero temperature is completely unlike QCD, but at finite temperature is in various respects analogous to
deconfined QCD.
Studies of energy loss have been conducted considering various types of partonic probes of the plasma, including quarks \cite{hkkky,gubser,ct,liu}, mesons \cite{sonnenschein,liu2,dragqqbar,dusling}, baryons \cite{draggluon,liu5,krishnan}, gluons \cite{draggluon,gubsergluon} and $k$-quarks \cite{draggluon}. For concreteness,
in what follows we will restrict attention to the case of a quark.

The rate of energy loss
determined analytically in the seminal works \cite{hkkky,gubser,ct} was only shown to apply for an isolated quark
 in the stationary or late-time regimes (i.e., when the quark moves
at constant velocity, under the influence of an external force which precisely cancels the drag force exerted by the plasma, or when the quark has
decelerated solely under the action of this drag force for a long period of time, and is about to come to rest). In the actual experimental setup, however, the configuration
is neither stationary nor asymptotic:
the quark is not externally forced, and slows down under the influence of the plasma, whose finite spatial and temporal extent imply that the
late-time regime is not generically accessible.\footnote{This latter point has been emphasized from the phenomenological perspective in \cite{peigne1}} Moreover, the real-world quark is not isolated, but is created within the plasma together with an antiquark.
In the present text we will explore to what extent the rate of energy loss is affected by these three issues.

\section{Basic Setup}

For our analysis we need two ingredients: a strongly-coupled thermal SYM plasma, and a quark that traverses it. The plasma is known to be described in dual language by a Schwarzschild black hole (actually, black threebrane) in asymptotically anti-de Sitter space, whose Hawking temperature is identified with $T$. More explicitly, to study $\mathcal{N}=4$ $SU(N_c)$
SYM with coupling $g_{YM}$ and temperature $T$, we must consider Type IIB string theory on the
(AdS-Schwarzschild)$_5\times\bS^5$ geometry
\begin{eqnarray}\label{metric}
ds^2&=&G_{mn}dx^m dx^n={R^2\over z^2}\left(
-hdt^2+d\vec{x}^2+{dz^2 \over h}\right)+R^2 d\Omega_5~, \\
h&=&1-\frac{z^{4}}{z_h^4}~, \qquad {R^4\over \ls^4}=g_{YM}^2 N_c\equiv\lambda~, \qquad
z_h={1\over \pi T}~,\nonumber
\end{eqnarray}
where $\ls$ denotes the string length, and $z_h$ is the location of the event horizon. In our work the  $\bS^5$ factor of this geometry
 will play no direct role. Notice that the field theory temperature is identified with the Hawking temperature $T_H=1/\pi z_h$ of the black hole.

 It is also known that a static heavy quark corresponds to a static and purely radial string extending down from
  the horizon at $z=z_h$ to a location $z=z_m$ where it ends on a  stack of $N_f$ D7-branes \cite{kk}. These branes cover the four gauge theory directions $t,\vec{x}$, and are spread
along the radial AdS direction from $z=0$ to $z=z_m$, where they `end' (meaning that the $\bS^3\subset\bS^5$
that they are wrapped on shrinks down to zero size). The D7-brane
parameter $z_m$ is related to the Lagrangian mass $m\gg \sqrt{\lambda}T$ of the
quark through \cite{hkkky}
\begin{equation}\label{zm} {1\over z_m}={2\pi
m\over\sqrt{\lambda}}\left[1+{1\over 8}\left(\sqrt{\lambda} T\over 2 m\right)^4-{5\over
128}\left(\sqrt{\lambda} T\over 2 m\right)^8+\cO\left(\left(\sqrt{\lambda} T\over 2
m\right)^{12}\right)\right]~.
\end{equation}
For applications of this formalism to phenomenology, we must choose values of the mass
parameter $z_m$ based on the charm and bottom quark masses, $m\simeq 1.4,4.8$ GeV. Following
\cite{gubsercompare}, this translates into $z_m/z_h\sim 0.16-0.40$ for charm and
$z_m/z_h\sim 0.046-0.11$ for bottom.

The string dynamics follows as usual from the Nambu-Goto action
\begin{equation}\label{nambugoto}
S_{\mbox{\scriptsize NG}}=-{1\over 2\pi\ap}\int
d^2\sigma\,\sqrt{-\det{g_{ab}}}\equiv {R^2\over 2\pi\ap}\int
d^2\sigma\,\cL_{\mbox{\scriptsize NG}}~,
\end{equation}
where $g_{ab}\equiv\partial_a X^m\partial_b X^n G_{mn}(X)$ ($a,b=0,1$) denotes
the induced metric on the worldsheet.

In the case of
vanishing temperature ($z_h\to\infty$), we are left
in (\ref{metric}) with a pure AdS geometry, and (\ref{zm}) reduces to
\begin{equation}\label{zmnoplasma}
z_m={\sqrt{\lambda}\over 2\pi m}~.
\end{equation}
One way to verify that the string is indeed dual to a quark is to compute the expectation value of the gluonic field in the presence of this object. The result is \cite{martinfsq}
\begin{equation}\label{Fsquaredq}
\expec{\tr
F^2(x)} ={\sqrt{\lambda}\over
16\pi^2|\vec{x}|^4}\left[1-\frac{1+{5\over 2}\left({2\pi
m|\vec{x}|\over\sqrt{\lambda}}\right)^2}{\left(1+\left({2\pi
m|\vec{x}|\over\sqrt{\lambda}}\right)^2\right)^{5/2}}\right]~.
\end{equation}
For $m\to\infty$ ($z_m\to 0$), this is just the Coulombic field expected (by conformal invariance) for a pointlike charge \cite{dkk}. For finite $m$ the profile is still Coulombic far away from the origin but in fact becomes non-singular at the location of the quark. We thus learn that the string is actually dual not to a pointlike `bare' quark, but to a `composite' or `dressed' quark, including its gluonic cloud. The size of this cloud
 is precisely the length scale $z_m$ defined in (\ref{zmnoplasma}), which is therefore understood to be the analog of the Compton wavelength for our non-Abelian source.

 Below we will also be interested in the description of a quark-antiquark pair. The IIB strings we have introduced above are oriented, and a state with two oppositely oriented purely radial strings would correspond to a quark and antiquark that are merely superposed. With such boundary conditions, however, the configuration with lowest energy is given by a \emph{single} $\cap$-shaped string that has both of its endpoints at $z=z_m$. This can again be verified by computing the corresponding gluonic profile. The result for infinitely massive quarks, at distances $|\vec{x}|$ much larger than the dipole separation $L$,  is \cite{cg}
 \begin{equation}\label{Fsquaredqqbar}
\expec{\tr
F^2(x)} ={15\Gamma\left(1\over 4\right)^4\sqrt{\lambda}L^3\over
256\pi^5|\vec{x}|^7}~,
\end{equation}
which has a more rapid falloff than (\ref{Fsquaredq}), as expected for an overall color-neutral source.\footnote{The additional factor of $L/|\vec{x}|$ in (\ref{Fsquaredqqbar}) beyond the usual electrostatic dipole behavior $L^2/|\vec{x}|^6$ is due to the large $N_c$ limit \cite{thorn}.}

 An important generic lesson of this type of computations is that the string codifies not only the quarks, but also the gluonic field generated by them. Roughly speaking, the would-be bare quarks/antiquarks are dual to the string endpoints, and the surrounding gluonic field is dual to the body of the string. In other words, we have learned through AdS/CFT that the usual `QCD' string exists even for non-confining theories, and actually lives in a curved 5 ($+$ 5)-dimensional geometry.

 Combining the two ingredients above, we know that a quark in a strongly-coupled SYM plasma is described by a string on the Schwarzschild-AdS geometry (\ref{metric}). The drag force felt by this quark when moving along a trajectory $\vec{x}(t)$ can then be deduced simply by having the string endpoint follow this same trajectory, and determining the pull of the body of the string which trails behind it. The works \cite{hkkky,gubser} derived an analytic formula for the corresponding rate of energy loss,
\begin{equation}\label{EPlossgubser}
\left({dE_q\over dt}\right)_{\mbox{\scriptsize s}}=-{\pi\over
2}\sqrt{\lambda}T^2\frac{v^2}{\sqrt{1-v^2}}~.
\end{equation}
As mentioned in the Introduction, this formula was proven to hold only in the stationary or late-time regimes, so \emph{a priori} it is not clear whether it could be relevant under conditions analogous to those of the experimental setup. This is the question that we now wish to examine. To orient ourselves, we will first inquire about the rate of energy loss for an accelerating quark in vacuum, whose dual description involves the string moving on pure AdS spacetime. In this case we will have full analytic control over the system, which will allow us to develop some intuition on the problem at hand. Additionally, our vacuum analysis will provide a useful benchmark against which the finite-temperature results can be compared.

\section{Isolated Quark at Zero Temperature}
\label{qnoplasmasec}

\subsection{Infinite mass}

\label{infinitemasssec}

A quark that accelerates in vacuum would be expected to emit chromoelectromagnetic
radiation. The first definite characterization of the radiation rate off an accelerating quark by means of AdS/CFT was
worked out in an important paper by Mikhailov \cite{mikhailov}. Remarkably, this author was able to solve the highly nonlinear
equation of motion for a string on AdS$_5$ that follows from (\ref{nambugoto}), for an \emph{arbitrary} timelike trajectory of
the string endpoint dual to a heavy quark! In terms of the coordinates used in
(\ref{metric}) (where for now $h=1$), his solution is
\begin{equation}\label{mikhsol}
X^{\mu}(\tau,z)=z{dx^{\mu}(\tau)\over d\tau}+x^{\mu}(\tau)~,
\end{equation}
with $\mu=0,1,2,3$, and $x^{\mu}(\tau)$ the worldline of the string endpoint at the AdS
boundary--- or, equivalently, the worldline of the dual, infinitely massive, quark---
parametrized by the proper time $\tau$ defined through
$\eta_{\mu\nu}\;\mathring{}\!\!x^{\mu}\;\mathring{}\!\!x^{\nu}=-1$, where
$\;\mathring{}\!\!x^{\mu}\equiv dx^{\mu}/d\tau$.

Combining (\ref{metric}) and (\ref{mikhsol}), the induced metric on the worldsheet is found
to be
$$
g_{\tau\tau}={R^2\over
z^2}(z^2\,\mathring{}\;\mathring{}\!\!\!x^2-1),\qquad
g_{zz}=0,\qquad g_{z\tau}=-{R^2\over z^2},
$$
implying in particular that the constant-$\tau$ lines are null, a fact that plays an
important role in Mikhailov's construction. In the solution (\ref{mikhsol}), the behavior at
time $t=X^{0}(\tau,z)$ of the string segment located at radial position $z$ (which according to our discussion in the previous section codifies the behavior of the gluonic field at the length scale $z$ \cite{uvir}) is completely
determined by the behavior of the  quark/endpoint at a retarded time $\tret(t,z)$ obtained
by projecting back toward the boundary along the null line at fixed $\tau$. {}From the
$\mu=0$ component of (\ref{mikhsol}), parametrizing the quark worldline by $x^0(\tau)$
instead of $\tau$, and using $d\tau=\sqrt{1-\vec{v}^{\,2}}dx^0$, where $\vec{v}\equiv
d\vec{x}/dx^0$, this amounts to
\begin{equation}\label{tret}
t=z{1\over\sqrt{1-\vec{v}^{\,2}}}+\tret~,
\end{equation}
where the endpoint velocity $\vec{v}$ is meant to be evaluated at $\tret$. In these same
terms, the spatial components of (\ref{mikhsol}) can be formulated as
\begin{equation}\label{xmikh}
\vec{X}(t,z)=z{\vec{v}\over\sqrt{1-\vec{v}^{\,2}}}+\vec{x}(\tret)=(t-\tret)\vec{v}+\vec{x}(\tret)~.
\end{equation}

Working in the static gauge $\sigma^0=t$, $\sigma^1=z$, the total energy of a string that
extends all the way down to the boundary--- i.e., with $z_m=0$, corresponding to an
infinitely massive quark--- follows from  the Nambu-Goto action (\ref{nambugoto}) as
\begin{equation}\label{estring}
E(t)={\sqrt{\lambda}\over 2\pi}\int_0^{\infty} {dz\over z^2}\frac{1+\left({\p\vec{X}\over\p
z}\right)^2} {\sqrt{1-\left({\p\vec{X}\over\p t}\right)^2+\left({\p\vec{X}\over\p
z}\right)^2-\left({\p\vec{X}\over\p t}\right)^2\left({\p\vec{X}\over\p
z}\right)^2+\left({\p\vec{X}\over\p t}\cdot{\p\vec{X}\over\p z}\right)^2}}~.
\end{equation}
Using (\ref{tret}) and (\ref{xmikh}), Mikhailov was able to reexpress this energy (via a
change of integration variable $z\to\tret$) as a local functional of the quark trajectory,
\begin{equation}\label{emikh}
E(t)={\sqrt{\lambda}\over 2\pi}\int^t_{-\infty}d\tret
\frac{\vec{a}^{\,2}-\left[\vec{v}\times\vec{a}\right]^2}{\left(1-\vec{v}^{\,2}\right)^3}
+E_q(\vec{v}(t))~,
\end{equation}
where of course $\vec{a}\equiv d\vec{v}/dx^0$. (A similar expression can be written down for the total string momentum.)
The second term in the above equation arises
from a total derivative that was not explicitly written down by Mikhailov, but can easily be
worked out to be \cite{chernicoff3}
\begin{equation}\label{edr}
E_q(\vec{v})={\sqrt{\lambda}\over
2\pi}\left.\left({1\over\sqrt{1-\vec{v}^{\,2}}}{1\over
z}\right)\right|^{z_m=0}_{\infty}=\gamma m~,
\end{equation}
which gives the expected Lorentz-invariant dispersion relation for
the quark. The energy split achieved in (\ref{emikh}) therefore
admits a clear and pleasant physical interpretation: $E_q$ (associated only
with information of the string endpoint) is the
intrinsic energy of the quark at time $t$, and the integral over
$\tret$ (associated with the body of the string) encodes the accumulated energy \emph{lost} by the quark to its gluonic field
over all times prior to $t$. No less remarkable is the fact that
the rate of energy loss for the quark in this strongly-coupled
non-Abelian theory is found to be in precise agreement with the
standard Lienard formula from classical electrodynamics!
The AdS/CFT correspondence thus teaches us that, in this very unfamiliar non-linear setting, the energy loss turns out to depend only locally on the quark worldline. This feature has been argued in \cite{kharzeev} to lead to an upper bound for the energy of a quark at finite temperature.

\subsection{Finite mass} \label{finitemasssec}

It is interesting to consider how these results are modified in the case $z_m>0$, where the
mass $m$ of the quark given by (\ref{zmnoplasma}) is large but not infinite.
In this case, we ought to impose boundary conditions on the string
not at $z=0$ but at $z=z_m$: given the worldline $\vec{x}(t)$ of the quark, we must require
that the string worldsheet satisfy $\vec{X}(t,z_m)=\vec{x}(t)$. Moreover, we need only
determine the behavior of the string in the region $z\ge z_m$.

Our task is to reexpress (\ref{mikhsol}) in terms of the data $\vec{x}(t)$ at the new
boundary $z=z_m$, instead of the (now merely auxiliary) data at
the AdS boundary $z=0$, which we will henceforth distinguish with
a tilde: $\vec{\tilde{x}}(t)$. For simplicity, we will carry out
this translation explicitly only in a setup where the quark moves
purely along direction $x\equiv x^1$, which is all that we will
need for our analysis in subsequent sections.\footnote{The corresponding formulas for arbitrary motion can be found in \cite{lorentzdirac} and a more detailed explanation in \cite{damping}.}

As was explained in \cite{chernicoff3}, it is possible to rewrite the
velocity $\tv$ and acceleration $\ta$ of the `auxiliary endpoint'
at $z=0$ in terms of the velocity $v\equiv dx/dt=\p_t X(t,z_m)$ of the actual string
endpoint and the momentum density $\Pi^z_x$ , which, when evaluated at $z=z_m$, controls the external force
$F=(\sqrt{\lambda}/2\pi)\Pi^z_x(t,z_m)$ acting on the string endpoint, or equivalently,
on the quark. One finds
\begin{eqnarray}\label{vtildeatilde}
\tv&=&\frac{v- z_m^2 \Pi }{1-z_m^2 v \Pi }~,\\
\ta&=&z_m \Pi \frac{(1-v^2)^{3/2}(1-z_m^4 \Pi^2)^{3/2}}{(1-z_m^2 v \Pi )^3}~,\nonumber
\end{eqnarray}
where we have abbreviated $\Pi\equiv\Pi^z_x$. Using (\ref{emikh}) (decorated with appropriate tildes), we can then express the total energy of the string at time $t$ in the form
\begin{equation}\label{emikhf}
E(t)=\int_{-\infty}^t \!dt\, \sqrt{\lambda}F^2\left({2\pi
m^2-\sqrt{\lambda} v F\over {4\pi^2 m^4-\lambda
F^2}}\right)+E_q(v(t),F(t))
~.
\end{equation}
As before, the first term represents the accumulated energy lost by the quark at all times
prior to $t$: it is the generalization to the $m<\infty$ case of the Lienard formula
(\ref{emikh}) deduced by Mikhailov (indeed, for $m\to\infty$ the integrand in (\ref{emikhf}) reduces to the Lienard expression for motion in one dimension, $(\sqrt{\lambda}/2\pi)F^2/m^2=(\sqrt{\lambda}/2\pi)a^2/(1-v^2)^{3}$).  The second term again denotes a surface term and gives
the modified dispersion relation for the finite-mass quark,
\begin{equation}\label{edrf}
E_q(v,F) ={\sqrt{\lambda}\over 2\pi}\left.\left({1-z_m^2 v \Pi
\over z\sqrt{(1-v^2)(1-z_m^4
\Pi^2)}}\right)\right|^{z_m}_{\infty}=\left({2\pi
m^2-\sqrt{\lambda} v F\over\sqrt{4\pi^2 m^4-\lambda
F^2}}\right)\gamma m~.
\end{equation}
Similar expressions can be written down for the momentum \cite{chernicoff3}.

Notice that (\ref{edrf}) differs from the result expected for a pointlike quark,
$E = \gamma m$. This reflects the fact that the fundamental source dual to a string that
terminates at $z_m>0$ is indeed \emph{not pointlike}. According to the standard dictionary mapping the radial coordinate in the curved geometry to a length scale in the gauge theory \cite{uvir}, it has a linear size of order $z_m$. It is only because of this
extended nature that, as we saw above, to characterize its state one needs to specify not
only the velocity $v$ but also the parameter $F$ (or $\p_z X(t,z_m)$) that encodes its
shape. The crucial point here is that, as we already noted below (\ref{Fsquaredq}), the source in question should not be thought of as a
bare quark, but as a `dressed' or `constituent' quark, surrounded by a gluonic cloud with
thickness $z_m$ \cite{martinfsq,mateos}.

Another salient feature of the energy of the quark given by the expression
(\ref{edrf}) is the fact that it diverges as the value of the
external force approaches
\begin{equation}\label{Fcritnoplasma}
F_{\mbox{\scriptsize crit}}= {2\pi m^2\over \sqrt{\lambda}}~.
\end{equation}
The reason for this is easy to understand on the string theory side. To exert a force $F$ on
the string endpoint, within the D7-branes we must turn on an electric field that has
strength $F_{01}=F$ at $z=z_m$.
The physical origin of the bound on the force is the fact that, for
$F_{01}>F^{\mbox{\scriptsize crit}}_{01}\equiv{\sqrt{\lambda}/
2\pi z_m^4}$, the creation of open
strings is energetically favored, and so the system is unstable.
According to (\ref{edrf}), then, the energy of the constituent quark diverges precisely at the point
where the external force becomes capable of nucleating
quark-antiquark pairs.

To summarize, from our study of a quark accelerating in vacuum we learn that: \emph{i)} the total energy of the string includes not only the intrinsic quark energy, but also the accumulated energy previously radiated by the quark; \emph{ii)} the quark dispersion relation arises from a surface term on the string worldsheet; \emph{iii)} the radiated energy depends localy on the quark worldline; \emph{iv)} when the quark has a finite mass, it is automatically non-pointlike, and there are consequently non-trivial modifications to its dispersion relation and radiation rate.

Before closing this section, we would like to make an additional observation. It is natural to expect the energy split achieved in (\ref{emikh}) or (\ref{emikhf}) to be somehow reflected in the geometry of the string worldsheet. It is interesting then that, on the string worldsheet dual to a quark undergoing arbitrary accelerated motion, there appears a black hole \cite{chernicoff3}, whose event horizon naturally divides the worldsheet into two causally disconnected regions (the appearance of a worldsheet black hole had also been noted previously at finite temperature \cite{gubserqhat,ctqhat}). It seems plausible then to interpret the regions outside and inside the black hole as corresponding respectively to the quark and the gluonic field, and postulate that the energy flow across this horizon should be related to the energy radiated by the quark \cite{dominguez,chernicoff3}. This possibility has been studied more closely in \cite{xiao,beuf,nothorizon}.

\section{Isolated Quark at Finite Temperature}
\label{qplasmasec}

\subsection{Constant velocity} \label{constantvsec}

Having understood the rate of energy loss for a heavy
quark that moves in the SYM vacuum, in this section we restore $z_h<\infty$--- and
consequently $h<1$--- in the metric (\ref{metric}), to study the same quantity in the case
where the quark moves through a thermal plasma.
A thorough generalization of Mikhailov's analytic results \cite{mikhailov} to this finite
temperature setup would require finding the exact solution to the Nambu-Goto equation of
motion for the string on the AdS-Schwarzschild background, for any given trajectory of the
string endpoint at $z_m\ge 0$. Sadly, we have not been able to accomplish this feat.
Nevertheless, based on the results discussed in the previous subsection, we expect the total
energy of the string at any given time to again decompose into a surface term that encodes
the intrinsic energy of the quark and an integrated local term that reflects the energy lost
by the quark.

There are two easy cases where one can show analytically that this expectation is borne out. The first is the case of a static quark, where there is of course no energy loss, and the string energy is correspondingly given by a pure surface contribution,
\begin{equation}\label{Mrest}
E={\sqrt{\lambda}\over
2\pi}\left({1\over z_m} - {1\over z_h}\right)\equiv M_{\mbox{\scriptsize rest}}~,
\end{equation}
which we naturally interpret as the rest mass of the quark in the thermal medium.
An important difference with respect to the $T=0$ case analyzed in
the previous subsection is that here the surface contribution
arises not only from the
lower ($z=z_m$) but also from the upper ($z=z_h$) endpoint of the
string. This is in fact the generic situation in the $T>0$ ($z_h<\infty$) case.
Since $z_h$ marks the position of an event horizon, for any finite coordinate time $t$ the value
of the surface contribution at $z_h$ is not influenced by the behavior of the $z<z_h$
portion of the string, but depends only on the string's configuration at $t\to-\infty$. The
same interpretation can be then carried over to the gauge theory: for arbitrary finite-temperature configurations, the terms in the quark dispersion relation arising from the upper string endpoint will encode a contribution to the energy of the state that depends solely on the initial
configuration of the quark$+$plasma system. Throughout the evolution, causality guarantees
that the behavior of the SYM fields at spatial infinity can only be affected by the initial
configuration at $t\to-\infty$, so we can equivalently think of the surface terms at $z_h$
as encoding information on the asymptotic boundary conditions for the system.  Indeed, for
dynamical processes, the radial location $z=z_h$ in AdS-Schwarzschild corresponds to the
deep IR of the gauge theory.

The only other
finite-temperature solution that is known analytically corresponds to a quark moving at constant velocity $v$, and takes the form
\cite{hkkky,gubser}
\begin{equation}\label{gubsersol}
X(t,z)=v\left[ t-{z_h\over 4}\ln\left(z_h+z\over z_h-z\right)+{z_h\over
2}\tan^{-1}\left({z\over z_h}\right)\right]~.
\end{equation}
 Imitating the procedure of the previous section, we can rewrite this solution by slicing it along null worldsheet curves,
\begin{equation}\label{gubsermikhsol}
X(\tret,z)={z_h v\over(1-v^2)^{1/4}}\tan^{-1}\left(z\over z_h(1-v^2)^{1/4}\right)+x(\tret)~,
\end{equation}
allowing the total energy of the string to be split into \cite{chernicoff3}
\begin{multline}\label{Emikhgubser}
E(t)=\frac{\sqrt{\lambda}}{2\pi }\int^{t}_{-\infty} d\tret\frac{v^2}{z_h^2\sqrt{1-v^2}}\\
+\frac{\sqrt{\lambda}}{2\pi}\left[
\frac{1}{z_m\sqrt{1-v^2}}+\frac{v^2}{z_h(1-v^2)^{\frac{3}{4}}}\tan^{-1}
\left(\frac{z_m}{z_h(1-v^2)^{\frac{1}{4}}}
\right)\right.\\
\left.-\frac{1}{z_h\sqrt{1-v^2}}-\frac{v^2}{z_h(1-v^2)^{\frac{3}{4}}}\tan^{-1}
\left(\frac{1}{(1-v^2)^{\frac{1}{4}}} \right) \right]~.
\end{multline}
As expected, the integrated term in the top line
recovers the result for the stationary rate of energy loss (\ref{EPlossgubser}).
The terms in the second and third line, then, codify the energy $E_q$ intrinsic to the
quark, including, as we discussed above, contributions dependent on the initial or boundary conditions. The total momentum of the string can be processed similarly.

\subsection{Accelerated motion} \label{acceleratedsec}

Having gained some intuition from the analysis of a quark moving
 at constant speed relative to the
strongly-coupled plasma, let us now turn our attention to the more
general situation where the quark accelerates. For concreteness,
 we will consider the early-time behavior of a quark
initially at rest, which is accelerated along a line by a time-dependent
external force that is turned off after a short interval,
allowing the quark to move thereafter only under the influence of
the plasma.

In \cite{chernicoff3} we integrated numerically
the Nambu-Goto equation of motion for the dual string.
Our results show a qualitative difference between the initial
stage ($0\le t < t_{\mbox{\scriptsize release}}$) where the quark
is accelerated by means of the external force $F(t)$, and the
second stage ($t_{\mbox{\scriptsize release}}\le t
<t_{\mbox{\scriptsize max}}$) where it moves only under the
influence of the plasma.
In the former stage,
the rate of energy loss, for values of $m$ in the neighborhood
of the charm or bottom masses, is in fact nearly identical
to the corresponding rate in vacuum, given by the modified Lienard formula
(\ref{emikhf}). In what follows we will elaborate a bit on the behavior
in the latter
stage, which would appear to be more relevant from the
phenomenological perspective.
A more detailed discussion of the energy lost by the quark
in both stages can be found in \cite{chernicoff3}.

\begin{figure}[tbph]
\vspace*{0.5cm}
 \setlength{\unitlength}{1cm}
\includegraphics[width=7cm,height=5cm]{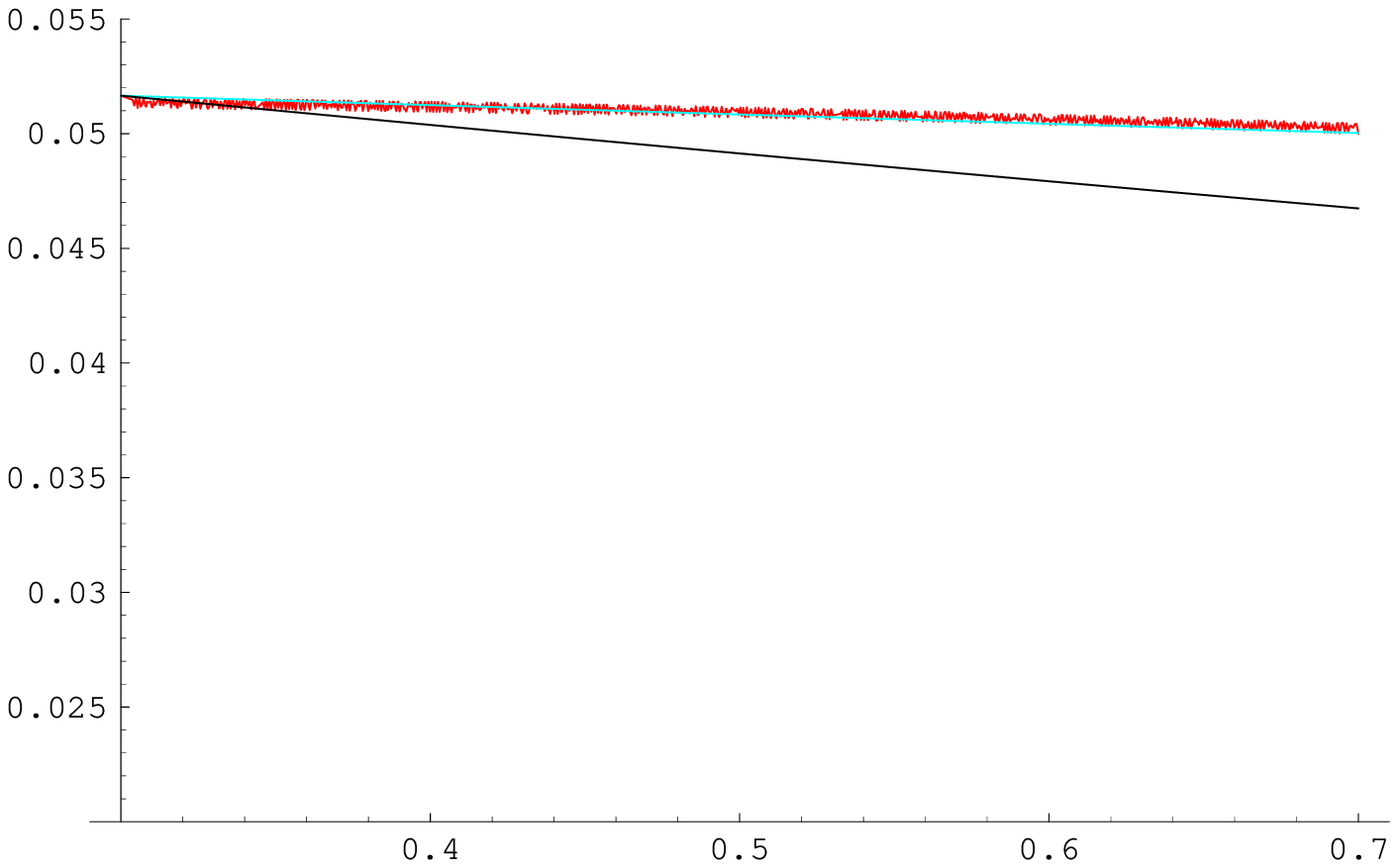}
 \begin{picture}(0,0)
   \put(0,0.3){$t$}
   \put(-6.5,5.1){$v$}
 \end{picture}
\includegraphics[width=7cm,height=5cm]{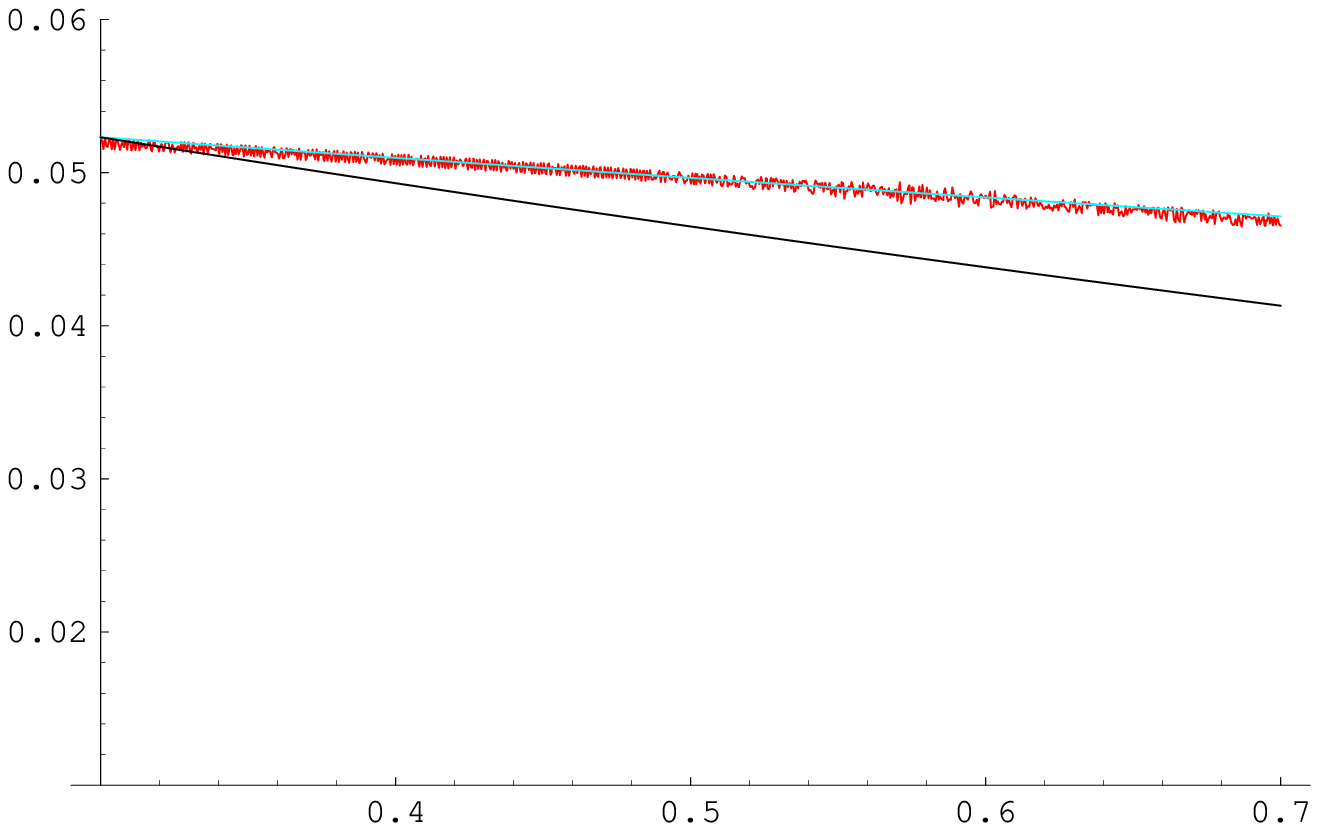}
 \begin{picture}(0,0)
   \put(0,0.3){$t$}
   \put(-6.5,5.1){$v$}
 \end{picture}
\caption{Quark velocity as a function of time (in units of 1/$\pi T$) from our numerical integration (in red)
compared against (\ref{vhkkkynr}) with the value of $\mu$ deduced in \cite{hkkky} (in
black), and (\ref{vhkkkynr}) with $\mu$ chosen to fit the data (in light blue), for a)
$z_m/z_h=0.2$ and b) $z_m/z_h=0.4$.}\label{vfig}
\end{figure}

The most direct way to inquire whether the output of our numerical integration for the
 accelerated quark conforms to the the constant-velocity (or late-time) results of  \cite{hkkky,gubser}
 is to compare the corresponding quark trajectories. If one assumes a dispersion
 relation $p_q=\gamma M_{\mbox{\scriptsize kin}} v$ as in \cite{hkkky} (which is what the vacuum expression (\ref{edrf}) predicts in the unforced case), then the rate of energy loss (\ref{EPlossgubser}) is equivalent to the statement that the friction coefficient
 \begin{equation}
 \mu\equiv -{1\over p_q}{dp_q\over dt}~.
 \end{equation}
 takes the constant value $\mu=\pi\sqrt{\lambda}T^2/2 M_{\mbox{\scriptsize kin}}$. The corresponding equation of motion for the quark is then
 \begin{equation}\label{qeomplasma}
{dv\over dt}=-\mu v(1-v^2)~,
 \end{equation}
 whose solution is \cite{hkkky}
 \begin{equation}\label{vhkkky}
v(t)=\frac{v_{\mbox{\scriptsize
release}}}{\sqrt{v_{\mbox{\scriptsize
release}}^2+(1-v_{\mbox{\scriptsize
release}}^2)e^{2\mu(t-t_{\mbox{\scriptsize release}})}}}~.
 \end{equation}
In our numerical results $v_{\mbox{\scriptsize release}}\ll 1$, so we are only able to test
the non-relativistic version of (\ref{vhkkky}),
\begin{equation}\label{vhkkkynr}
v(t)=v_{\mbox{\scriptsize release}}e^{-\mu(t-t_{\mbox{\scriptsize
release}})}~
\end{equation} (and, since our integration is limited to small time intervals, we
would in effect see just the linear portion of this function). A
comparison between this analytic prediction and our numerical
results for $t\ge t_{\mbox{\scriptsize release}}$ is given in
Fig.~\ref{vfig}. It is evident from this plot that, in the early
stage of motion covered by our analysis, the quark dissipates
energy at a rate much lower than the late-time result of
\cite{hkkky}. Indeed, for $z_m/z_h=0.2,0.3,0.4$ the late-time
friction coefficient is respectively $\mu_{\mbox{\scriptsize
late}}/\pi T=0.25,0.41,0.59$, but our numerical results for $v(t)$
are best approximated by $\mu_{\mbox{\scriptsize early}}/\pi
T=0.08,0.15,0.26$.

The relation between energy loss and the appearance of a black hole on the worldsheet in this finite temperature setting had been noted already in \cite{gubserqhat,ctqhat} for the case of a stationary configuration. The non-stationary case and the parallel with the zero temperature situation were discussed in \cite{chernicoff3,xiao,beuf}.

An interesting prediction of AdS/CFT is the existence of a subluminal limiting velocity $v_m\equiv \sqrt{1-z_m^4/z_h^4}\simeq 1-(\sqrt{\lambda}T/2m)^4$ for the quark traversing the plasma. This follows simply from the fact that the quark velocity is dual to the \emph{coordinate} velocity of the string endpoint, and $v=v_m$ corresponds to a \emph{proper} velocity at $z=z_m$ equal to that of light \cite{argyres1}. A more general bound involving the external force $\vec{F}$ can be derived by requiring the Nambu-Goto square root to remain real \cite{chernicoff3}, which reduces to $v\le v_m$ when $\vec{F}=0$. This same restriction on the velocity can be seen to arise from microscopic calculations of the meson spectrum \cite{mateosthermo,liu4}. Its validity for isolated quarks has been emphasized in \cite{argyres3}.

\section{Pair Creation within the Plasma}
\label{qqbarsec}

As was mentioned in the Introduction, in the experimental setup the heavy quark is created within the plasma together with its corresponding antiquark, and the presence of the latter would be expected to substantially modify the gluonic fields in the vicinity of the quark, consequently affecting its evolution. In \cite{chernicoff3}, building upon the previous work \cite{hkkky}, we have numerically studied the evolution of
a heavy quark and antiquark that are created within the plasma at time $t=0$ and
then separate back to back. In dual language, this corresponds to a string with both of its endpoints on the D7-branes at $z=z_m$, such that the endpoints are initially at the same spatial location but have initial velocities in opposite directions.

In setting up the problem, one realizes that on the gauge theory side there are actually two distinct quark-antiquark configurations:  the product of a fundamental $q$ and an antifundamental $\bar{q}$ can lead
to a $q$-$\bar{q}$ pair either in the singlet or the adjoint representation of the $SU(N_c)$
gauge group. Interestingly, we find that there is a counterpart to this in the gravity side, because there are precisely two distinct types of consistent initial conditions for the string.  The  most obvious possibility is a string that is
completely pointlike at $t=0$ \cite{hkkky}, and then expands into a $\cap$-shape as its endpoints separate.
Through the standard recipe for correlation functions \cite{gkpw}, it is clear that, initially, this configuration sets up no long-range chromoelectromagnetic field, and so
describes the singlet. The only other allowed possibility is a string that is initially
 linelike \cite{chernicoff3}, extending from $z_m$ all the way up to $z_h$, and expands for $t>0$ into a $\wedge$-shape. Such linelike string is precisely the system considered in
\cite{gubserstable,draggluon} to model a color source in the adjoint
representation. Due to its extended nature, it sets up long-range supergravity fields that
translate into a long-range gluonic field
profile, indicative of a source with net color charge.
Curiously, we find that the initial quark velocity is freely
adjustable in the adjoint, but not the singlet, configuration--- in the latter case it is invariably fixed at the limiting value $v=v_m$.

\begin{figure}[htb]
\begin{center}
\vspace*{0.2cm} \setlength{\unitlength}{1cm}
\includegraphics[width=9cm,height=6cm]{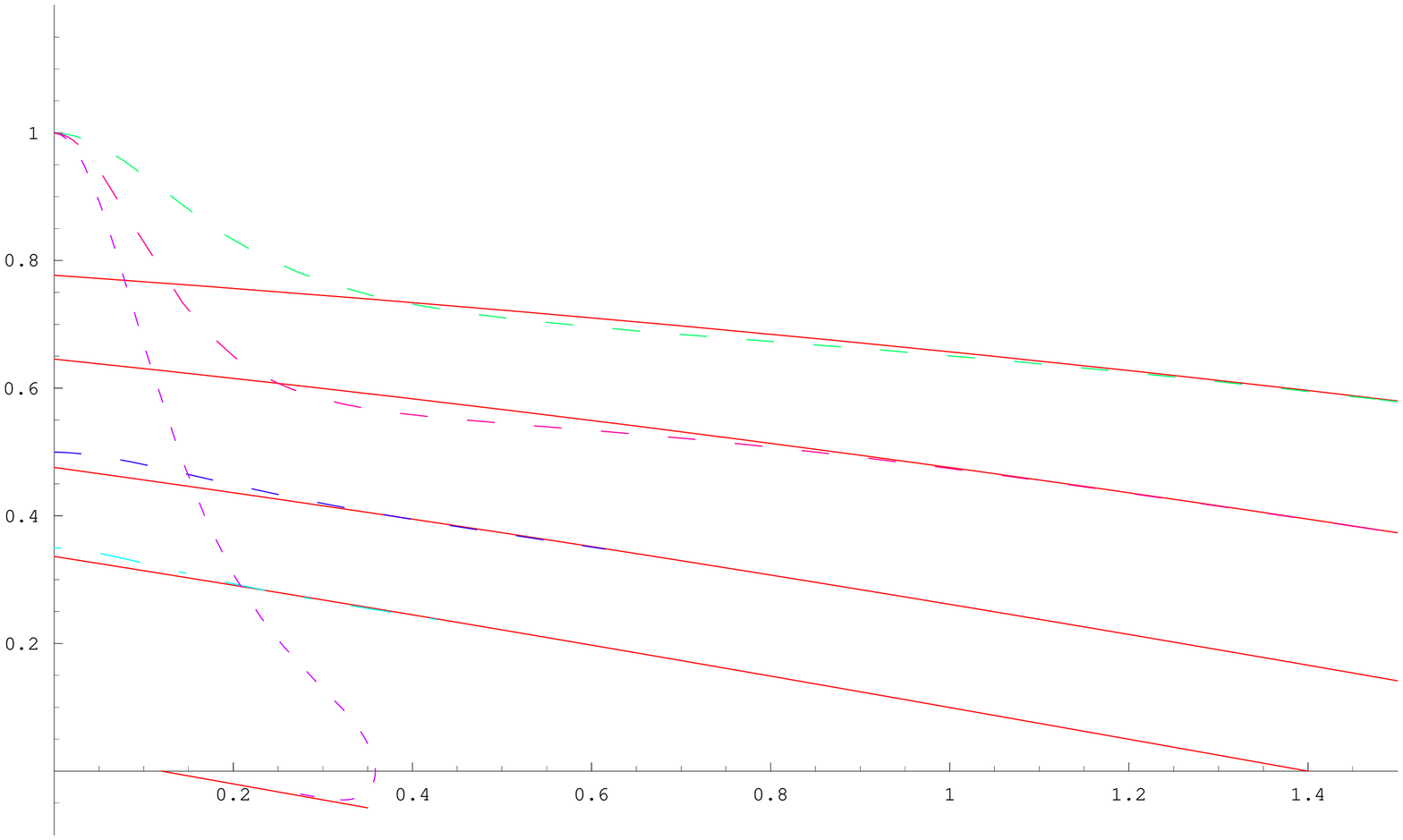}
 \begin{picture}(0,0)
   \put(0.1,0.4){$x$}
   \put(-8.7,6.1){$v$}
   \put(-6.8,5){\vector(0,-1){1}}
    \put(-4.9,4.0){\vector(0,-1){1}}
   \put(-6.8,1.4){\vector(0,-1){1}}
  \put(-8.7,3.8){\vector(0,-1){1}}
   \put(-8.7,1.0){\vector(0,1){1}}
    \end{picture}\hspace{1cm}
\caption{Quark evolution (velocity as a function of traveled distance,
in units of $1/\pi T$)
 for five different initial conditions. The dotted curves show the results
 of our numerical integration, contrasted against fits in solid red that use
  the analytic expression (\ref{constmuvx}), with the value $\mu=0.25$ obtained
  in \cite{hkkky} and an optimal choice of the stopping distance
  $x_{\infty}$. The three dotted curves starting
  at $v_0=v_m$ describe singlet configurations with different
  initial velocity profiles and energies.
Notice in particular that the purple curve describes a
situation where the quark and antiquarks turn around and come to rest while approaching
one another. The two remaining curves arise from adjoint configurations with different energies and initial quark
velocities. The vertical arrows mark the transition distance where each trajectory enters the late-time regime described by (\ref{constmuvx}).} \label{stage2fig}
\end{center}
\end{figure}

As the quark and antiquark separate, we expect them to eventually enter the late-time regime where their rate of energy loss is given by (\ref{EPlossgubser}) and their trajectory is given by (\ref{vhkkky}) (with $t_{\mbox{\scriptsize release}}$ and
$v_{\mbox{\scriptsize release}}$ regarded as adjustable parameters).
The agreement between the analytic and late-time numeric results is
most cleanly seen if instead of comparing graphs of $x(t)$ or
$v(t)$ for the quark (where one would need to look at
$t\to\infty$), one examines the plots of $v(x)$, where the analytic
behavior for constant $\mu$ takes the simple form
\begin{equation}\label{constmuvx}
v(x)=\tanh[\mu(x_{\infty} - x)]~,
\end{equation}
which is linear with slope $-\mu$ near the final rest point
$x=x_{\infty}$ (whose value is meant to be adjusted to fit the
data). These plots are shown in Fig.~\ref{stage2fig}, for
  mass parameter $z_m/z_h=0.2$ (in the neighborhood
 of the charm quark), corresponding
  to a limiting velocity $v_m=0.9992$. It is
evident from the figure that the late-time behavior is well-described by
(\ref{constmuvx}) in all cases.

In Fig.~\ref{stage2fig}, we see that there is an initial period
where the behavior differs
 from the late-time frictional evolution (\ref{constmuvx}).
 This difference is clearly more significant for
 the singlet than the adjoint case. For values of the mass in the neighborhood of the charm or bottom masses, our numerical results indicate that the initial evolution is essentially identical to what it would be in the absence of the plasma.
  This picture seems rather close to the phenomenological
  discussion given
  in \cite{peigne1} (in the context of collisional
 energy loss):
  when the singlet quark-antiquark pair is formed
 within the plasma, there is a delay before
 the interaction between the newly created sources and the plasma
 can set up the long range gluonic field profile that is responsible
 for the late-time dissipation.
To examine in more detail the transition to the late-time behavior
(\ref{constmuvx}),  for a variety of trajectories we can
determine the point $(x_f,v_f)$ beyond which the numeric $v(x)$
curve agrees with the analytic curve (\ref{constmuvx}) to a given
accuracy $f$. (In \cite{chernicoff3} we considered $f=5$ or 10\%.)
The results
are schematically
indicated by the arrows in Fig.~\ref{stage2fig}.

 By the time when the quark moves beyond $x_f(v_f)$ and therefore
 enters the late-time regime, it is certainly insensitive to the presence of the antiquark. It is natural then to ask how the transition distance $x_f(v_f)$ compares against (half of) the length $L_{\mbox{\scriptsize max}}(v)$ beyond which the quark and antiquark are screened from each other by the plasma. Based on what we just said, we know that for a given velocity we must have $L_{\mbox{\scriptsize max}}/2 <x_f$, but the actual comparison will tell us  whether the transition to the regime
 where the quark experiences a constant drag coefficient occurs
 right after the quark and antiquark are screened from each other
 by the plasma, or if there is an intermediate regime where the
 quark moves independently from the antiquark but nevertheless
 feels a drag force that differs from the stationary result of \cite{hkkky,gubser,ct},
 as we found when applying an external force to the isolated quark in Section
 \ref{acceleratedsec}.

\begin{figure}[htb]
\vspace*{0.5cm}
\begin{center}
\setlength{\unitlength}{1cm}
\includegraphics[width=6cm,height=4cm]{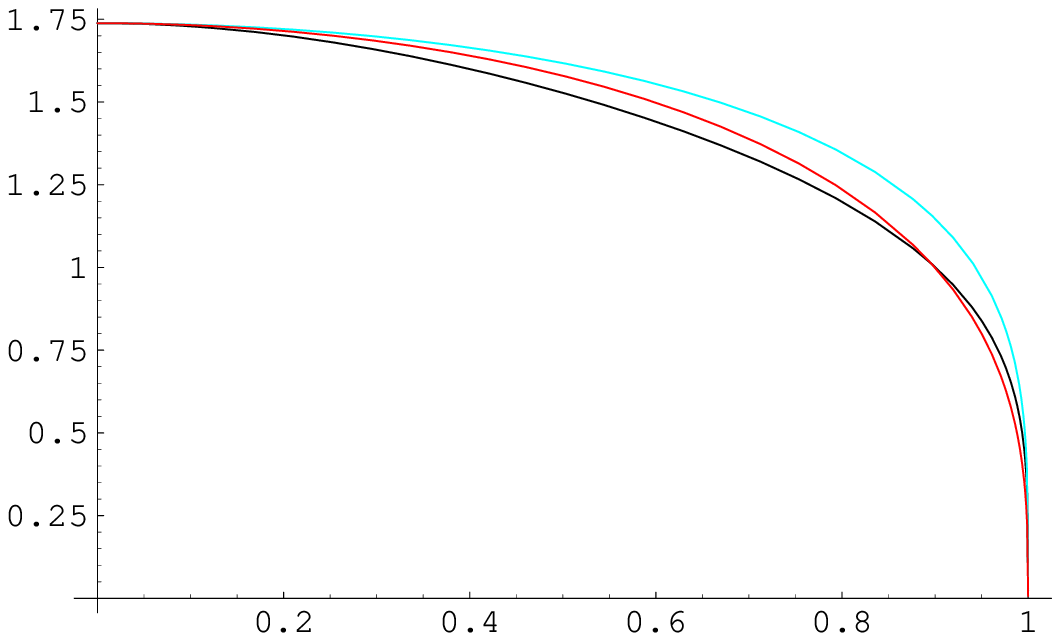}
 \begin{picture}(0,0)
   \put(0.2,0.3){$v$}
   \put(-6.1 ,4.2){$L_{\mbox{\scriptsize max}}$}
 \end{picture}
\caption{Screening length $L_{\mbox{\scriptsize max}}$ (in units
of $1/2\pi T$) as a
function of velocity (in black) compared against the
approximations (\ref{onethird}) (in red) and (\ref{onequarter}) (in
blue).}\label{onethirdfig}
\end{center}
\end{figure}

The screening length for infinitely massive quarks in $\cN=4$ SYM was computed in \cite{liu2,dragqqbar}, by considering a quark-antiquark pair moving jointly through the plasma (a related length was plotted in
 \cite{sonnenschein}). Over
the entire range $0\le v\le 1$ its behavior may be approximated
as \cite{dragqqbar}
\begin{equation} \label{onethird}
L_{\mbox{\scriptsize max}}(v)\approx{0.865\over \pi
T}(1-v^2)^{1/3}~,
\end{equation}
while in the ultra-relativistic limit, it can be shown analytically that \cite{liu2}
\begin{equation}\label{onequarteranal}
L_{\mbox{\scriptsize max}}(v)\to {1\over \pi T}{
3^{-3/4}4\pi^{3/2}\over\Gamma(1/4)^2}(1-v^2)^{1/4}\simeq
{0.743\over \pi T}(1-v^2)^{1/4}\quad\mbox{for $v\to 1$.}
\end{equation}
The full curve $L_{\mbox{\scriptsize max}}(v)$ does not deviate
far from this asymptotic form, so a decent
approximation to it is obtained by replacing $0.743\to 0.865$ in
(\ref{onequarter}), to reproduce the correct value at $v=0$ (at
the expense of introducing a 16\% error as $v\to 1$) \cite{liu2}:
\begin{equation} \label{onequarter}
L_{\mbox{\scriptsize max}}(v)\approx{0.865\over \pi
T}(1-v^2)^{1/4}~.
\end{equation}
A comparison between the two approximations (\ref{onethird})
 and (\ref{onequarter}) is shown
 in Fig.~\ref{onethirdfig}: overall, the
 exponent $1/3$
is better than $1/4$ in the sense that
it implies a smaller squared deviation from the numerical results,
even though $1/4$ leads
to a smaller \emph{percentage} error in the range $v>0.991$
($\gamma>7.3$).
An attempt to better parametrize the deviation away from the
ultra-relativistic behavior was made in \cite{sfetsosqqbar}.
In any case, one should bear in mind that the region of principal
interest at RHIC is not really $v\to 1$, but
 $\gamma v\sim 1$.

\begin{figure}[htb]
 \vspace*{0.5cm}
\setlength{\unitlength}{1cm}
\includegraphics[width=6cm,height=4cm]{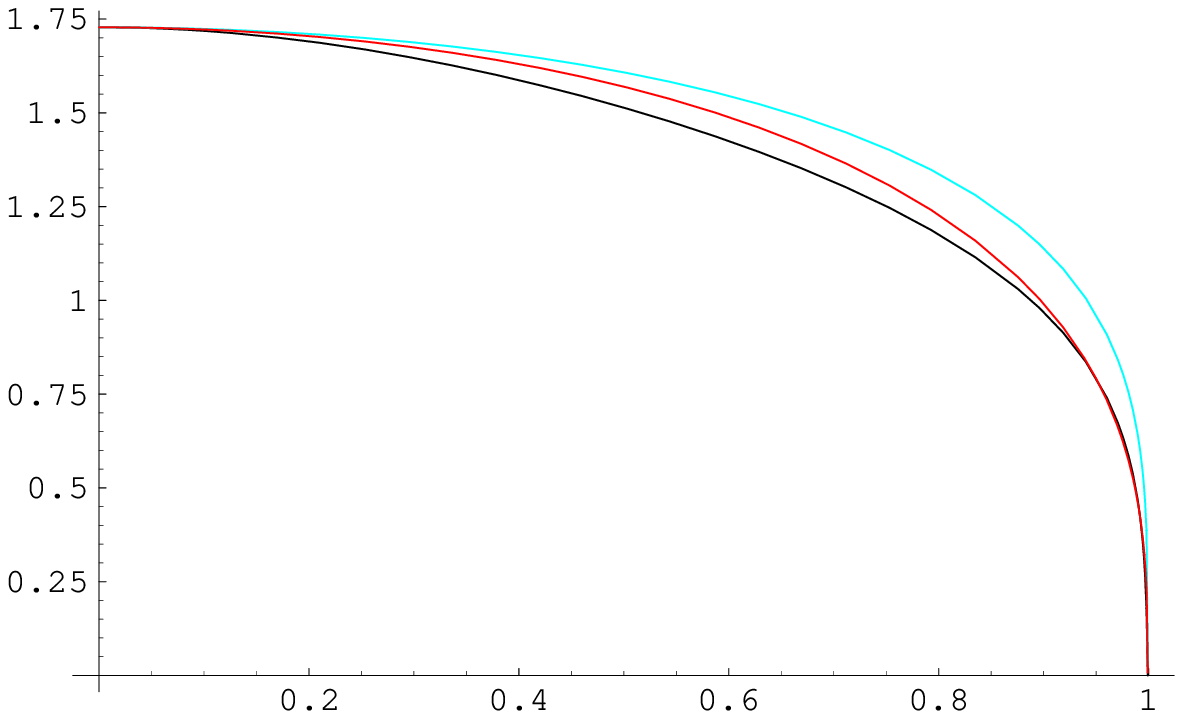}
 \begin{picture}(0,0)
   \put(0.2,0.2){$v$}
   \put(-5.6,4.1){$L_{\mbox{\scriptsize max}}$}
 \end{picture}\hspace{1cm}
 \includegraphics[width=6cm,height=4cm]{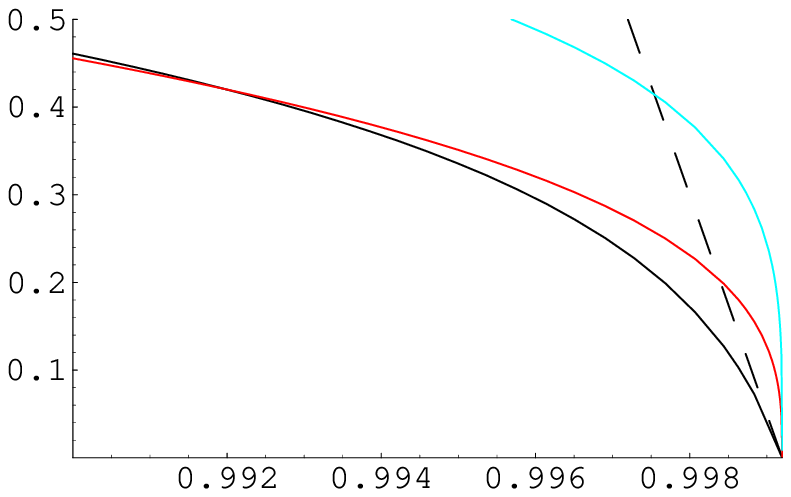}
 \begin{picture}(0,0)
   \put(0.2,0.2){$v$}
   \put(-5.6,4.1){$L_{\mbox{\scriptsize max}}$}
 \end{picture}
\caption{(a) Screening length $L_{\mbox{\scriptsize max}}$ (in
units of $2\pi T$) as a function of velocity for $z_m/z_h=0.2$ (in
black) compared against the $z_m>0$ fits (\ref{onethirdvm}) (in red)
and (\ref{onequartervm}) (in light blue). (b) Expanded version of
the same plot, showing that (\ref{onethirdvm}) gives a relatively
good approximation up to velocities very close to $v_m$, where the
asymptotic linear behavior (\ref{one}) (dotted dark blue) sets
in.} \label{lmaxfig2}
\end{figure}

To compare against the transition distance $x_f$ defined above, in \cite{chernicoff3} we extended this calculation to the case of finite mass. Fig.~\ref{lmaxfig2} shows the result for a mass in the neighborhood of the charm value. The main difference with respect to the $m\to\infty$ case is
 that now the `ultra-relativistic' region would refer to the limit where the pair velocity
 approaches the limiting velocity $v_m<1$.
The screening length
$L_{\mbox{\scriptsize max}}(v)$ is still relatively well
approximated in the full range $0\le v\le v_m$ by the natural modification
of
the $z_m=0$
fit (\ref{onethird}),
\begin{equation} \label{onethirdvm}
L_{\mbox{\scriptsize max}}(v)\approx{0.865\over \pi
T v_m^{2/3}}(v_m^2-v^2)^{1/3}~.
\end{equation}
This approximation becomes worse as $z_m/z_h$
is further increased. (In all cases, the fit analogous
to (\ref{onequarter}),
\begin{equation}\label{onequartervm}
L_{\mbox{\scriptsize max}}(v)\approx
{0.865\over \pi T v_m^{1/2}}(v_m^2-v^2)^{1/4}~,
\end{equation}
 does a poorer job than
(\ref{onethirdvm}).)
The behavior in the $v\to v_m$ region
 can still be determined analytically, and turns out to be
 \begin{equation}\label{one}
L_{\mbox{\scriptsize max}}\to {1\over 2\pi T}{v_m^2-v^2\over
v_m(1-v_m^2)^{3/4}}={1\over 2\pi T}{z_h^3[1-(z_m/z_h)^4-v^2]\over
z_m^3\sqrt{1-(z_m/z_h)^4}}~,
 \end{equation}
 where we see that the $1/4$ exponent in (\ref{onequarteranal}) changes to $1$ at finite quark
 mass.

 \begin{figure}[htb]
 \vspace{0.2cm}
 \begin{center}
\setlength{\unitlength}{1cm}
\includegraphics[width=6cm,height=4cm]{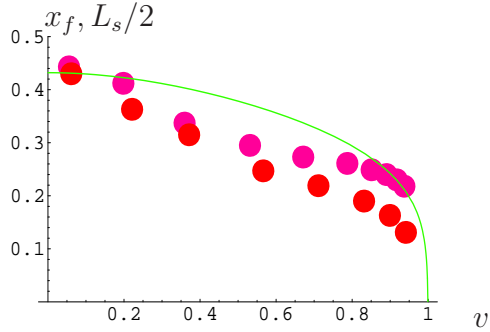}
 \begin{picture}(0,0)
   \put(0.2,0){$v$}
   \put(-5.5,4){$x_{f},L_{s}/2$}
 \end{picture}\hspace{1cm}
 \caption{Transition distance $x_f(v)$, with $z_m/z_h=0.2$, for $f=0.1$ (red) and
$f=0.05$ (magenta), compared against (half of) the screening length $L_{\mbox{\scriptsize max}}(v)$ (green). The vertical axis is in units of $1/\pi T$.} \label{xlmaxfig}
\end{center}
\end{figure}

We can now finally perform the promised comparison between the transition distance $x_f(v)$ and (half of) the screening length $L_{\mbox{\scriptsize max}}(v)$. The result,
for singlet configurations, is shown in Fig.~\ref{xlmaxfig},
where we see that the two separations are of comparable magnitude and scale with velocity in
a similar manner.\footnote{For adjoint configurations, the two lengths also agree but have a trivial $v$ depencence: $x_f$ is seen in Fig.~\ref{stage2fig} to be essentially equal to zero, and the adjoint quark-antiquark potential is suppressed by a factor of $1/N_c^2$
(see, e.g., \cite{adjointpot}), implying that $L_{\mbox{\scriptsize max}}=0$ at large $N_c$.} A similar agreement was reported in \cite{iancu2} (later related work may be found in \cite{liustirring}). Notice that this is in spite of the fact that the two relevant string
configurations are quite different, with motion respectively along and perpendicular to the
plane in which the string extends.

 We conclude then that the transition to the
constant-drag-coefficient regime takes place immediately after the quark and antiquark lose
contact with one another. That is to say, unlike what we found for the forced isolated quark
in Section \ref{acceleratedsec}, here there is no intermediate stage where the quark and
antiquark decelerate independently from one another at a rate that differs substantially
from the late-time result of \cite{hkkky,gubser}.

To summarize, in the more realistic scenario where the quark and antiquark are created within the plasma and then separate back-to-back, we find that there are two distinct stages of the evolution. At the beginning, the quark essentially does not feel the plasma and slows down only under the influence of the antiquark, just as it would in vacuum. To the extent that this result, with all its simplifying assumptions,
might conceivably be extrapolated
to the experimental context,
this initial stage would not differ significantly
between heavy ion and proton-proton collisions. Later, there comes a point when
the  $q$-$\bar{q}$ separation $L_{q\bar{q}}$ becomes large enough, and the quark velocity low enough, that the pair crosses the relevant screening length, i.e., $L_{q\bar{q}}\approx L_{\mbox{\scriptsize max}}(v)$. Beyond this point the quark starts feeling the plasma but no longer feels the antiquark, and evolves exactly as the  expressions (\ref{EPlossgubser}) and (\ref{vhkkky}) predict. This then extends and at the same time delimits the region
where the analytic results of \cite{hkkky,gubser,ct} can justifiably be used to model energy loss in heavy ion collisions.

\section*{Acknowledgements}
We wish to thank Antonio Garc\'\i a for useful conversations and collaboration in the initial stages of this work. We are also grateful to the organizers of the XIII Mexican School of Particles and Fields for the invitation and for putting together a very productive meeting. This work was
partially supported by Mexico's National Council of Science and Technology (CONACyT) grant
50-155I, as well as by DGAPA-UNAM grant IN116408.

\end{document}